\documentclass[a4paper,letter]{aa}

\usepackage{graphicx}
\usepackage{txfonts}
\usepackage{natbib}
\usepackage{lscape}
\usepackage[dvipdfm,breaklinks=true,bookmarks=true,hypertexnames=false,bookmarksnumbered=true,bookmarksopen=false,bookmarks=true,bookmarksnumbered=true]{hyperref}

\newcommand{\msun}{{\rm M}_{\odot}}

\newcommand{\kms}{\, {\rm km\, s}^{-1}}
\newcommand{\h}{\,h_{70}}
\newcommand{\hmm}{\h^{-1}}
\newcommand{\hmmsun}{\hmm\msun}
\newcommand{\kpc}{\, {\rm kpc}}

\newcommand{\hmkpc}{\hmm\kpc}

\newcommand{\Mpc}{\, {\rm Mpc}}
\newcommand{\hmMpc}{\hmm\Mpc}
\newcommand{\der}{{\rm d}}

\newcommand{\nbg}{n_{\rm bg}}
\newcommand{\scrit}{\Sigma_{\rm crit}}

\newcommand{\eg}{{\it e.g.}}
\newcommand{\ie}{{\it i.e.}}

\newcommand{\zp}{z_{\rm phot}}

\newcommand{\mypm}[2]{^{+#1}_{-#2}}

\begin{document}

\title{A weak lensing study of the Coma Cluster
\thanks{Based on observations obtained with MegaPrime/MegaCam, a joint project
of CFHT and CEA/DAPNIA, at the Canada-France-Hawaii Telescope (CFHT) which
is operated by the National Research Council (NRC) of Canada, the Institut
National des Sciences de l'Univers of the Centre National de la Recherche
Scientifique (CNRS) of France, and the University of Hawaii. This work is
 also partly based on data products produced at TERAPIX and the Canadian
Astronomy Data Centre as part of the Canada-France-Hawaii Telescope Legacy
Survey, a collaborative project of NRC and CNRS.}}

\author{
  R. Gavazzi\inst{1},
  C. Adami\inst{2},
  F. Durret\inst{1},
  J.-C. Cuillandre\inst{3},
  O. Ilbert\inst{2},
  A. Mazure\inst{2},
  R. Pell\'o\inst{4},
  M.P. Ulmer\inst{5}
}

\institute{
  Institut d'Astrophysique de Paris, 
  CNRS UMR7095 \& Univ. Paris 6, 98bis Bd Arago, 75014 Paris, France
  \and
  LAM, OAMP, Universit\'e Aix-Marseille $\&$ CNRS, 38 rue Fr\'ed\'eric Joliot-Curie,
  13388 Marseille 13 Cedex, France
\and
Canada-France-Hawaii Telescope Corporation, Kamuela, HI 96743, USA
\and
Laboratoire d'Astrophysique de Toulouse-Tarbes, Universit\'e de Toulouse,
CNRS, 14 Av. Edouard Belin, 
31400 Toulouse, France
\and
Department Physics $\&$ Astronomy, Northwestern University, Evanston, IL 60208-2900, USA
}


\abstract
{Due to observational constraints, dark matter determinations in nearby
clusters based on weak lensing are still extremely rare, in spite of
their importance for the determination of cluster properties independent of
other methods.
}
{We present a weak lensing study of the Coma cluster (redshift$\sim$
0.024) based on deep images obtained at the CFHT.
}
{After obtaining photometric redshifts for the galaxies in our
  field based on deep images in the $u^*$ (1$\times$1 deg$^2$), and in the B, V, R
  and I bands (42'$\times$52'), allowing us to eliminate foreground galaxies, 
  we apply weak lensing calculations on shape measurements performed in
  the  $u^*$ image.
}
{We derive a map of the mass distribution
in Coma, as well as the radial shear profile, and the mass and concentration
parameter at various radii.
We obtain $M_{\rm 200c}= 5.1\mypm{4.3}{2.1} \times 10^{14} \hmmsun$ and
$c_{\rm 200c}=5.0\mypm{3.2}{2.5}$, in good agreement with previous measurements.}
{With deep wide field images it is now possible to analyze nearby clusters with
  weak lensing techniques, thus opening a broad new field of investigation.
}


\authorrunning{Gavazzi et al.}
\titlerunning{Weak lensing in Coma}

\maketitle


\section{Introduction}


Although weak lensing studies are of great importance in cosmology, in
particular concerning various aspects of the analysis of large scale
structures in the universe \citep[\eg~][]{mellier99,BS01,schneider06rev},
dark matter
determinations in clusters based on weak lensing have been limited to
relatively distant objects, usually with redshifts larger than about
0.15. This is mainly due to the fact that the cluster lensing strength
is maximized if the angular diameter distance to the foreground cluster
is about half that of the background source galaxies, and to the
requirement of large field images, in order to cover the full cluster in
a single exposure. Very few low redshift clusters have been analyzed 
until now in detail through weak lensing: Abell~3667 at redshift 0.055
\citep{joffre00}, and clusters observed in the shallow
Sloan Digital Sky Survey \citep{sheldon01}, including Coma \citep{kubo07}. 
We present here a weak lensing study of the Coma cluster at
$z_{\rm Coma}=0.024$ based on 
deep exposures in several bands obtained with the CFH12k and
Megaprime/Megacam large cameras at the Canada France Hawaii Telescope.
The J2000 location of the brightest cluster galaxy NGC~4874 is 
RA=12:59:35.67, DEC=+27:57:33.7\ .
In the following we assume the {\it ``concordance model''}
cosmological background with $H_0 = 70\,\kms\Mpc^{-1}$,
$\Omega_{\rm m}=0.3$ and $\Omega_\Lambda=0.7$, for which 
$1\arcsec$ on the sky corresponds to a transverse scale of $0.79\kpc$.
All magnitudes are in the AB system.

\section{Data analysis}\label{sec:method}


\subsection{Cluster membership \& background sources}\label{ssec:members}
Multiband imaging for this cluster was obtained at CFHT, both in the
inner parts made of two CFH12k camera pointings in BVRI bands resulting
in about 0.54 deg$^2$ coverage and with Megacam ($1\times1\, {\rm
deg}^2$ fov) in the $u^*$ band \citep[see][ for a detailed description
of these data]{adami06,adami07,adami08}.

We were able to estimate photometric redshifts of all sources down to 
$R\lesssim 23$ in the area covered by the CFH12k images
\citep{adami08}.  This allowed us to separate the population of faint
background sources from nearby objects with $\zp\le0.04$ and 
to eliminate the bulk of the cluster members that could bias
the weak lensing study \citep[\eg~][]{medezinski07}.
\citet{adami08} used the same
technique to discriminate between $z<0.2$ and $z>0.2$ galaxies. Here, we
push the limit to $\zp=0.04$ to increase the number of
background sources although a more conservative cut at $0.2$ would not
produce noticeable changes given the small number of added objects.

This photometric redshift separation is possible only in the innermost
0.54 deg$^2$. However we found that none of our results below could be
significantly biased due to the extreme prevalence of distant sources in
the faint end of the magnitude counts (a factor $\sim 10-20$) despite some
incompleteness for $u^*\gtrsim 24.5$\ .
We thus apply a magnitude selection $23.5 \le u^*\le25$ everywhere in the
field but we complement the selection by removing the few objects in the inner
regions that are known to be at $\zp\le0.04$.

A clear advantage of lensing by nearby deflectors is that the geometric
distance factor $D_{\rm LS}/D_{\rm S}$ \citep[\eg~][]{BS01}, which is the
ratio of angular distances between deflector and source $D_{\rm LS}$ and
to the source $D_{\rm S}$, is almost
constant and close to unity for most of the faint background
sources. Given the low redshift of Coma, even a source
at $z\sim0.2$ will experience lensing effects with $D_{\rm LS}/D_{\rm
S}\simeq 0.87$. In order to check this small redshift dependance,
we used the CFHTLS-deep
photometric redshifts \citep{ilbert06} measured in four independent 1
deg$^2$ fields to infer the redshift distribution of sources having
$23.5 \le u^* \le 25$. The mean and median photometric redshifts are
0.87 and 0.92 respectively and the sample contains a negligible amount
of low redshift ($z<0.04$) galaxies. We estimate the mean
$\overline{w}=\langle D_{\rm LS}/D_{\rm S}\rangle$ averaged over the
whole population of sources to be $\overline{w}\simeq 0.962$.  This is
the value we will consider below for further weak lensing mass
calibrations.

\begin{figure}[htb]
  \centering \includegraphics[width=\hsize]{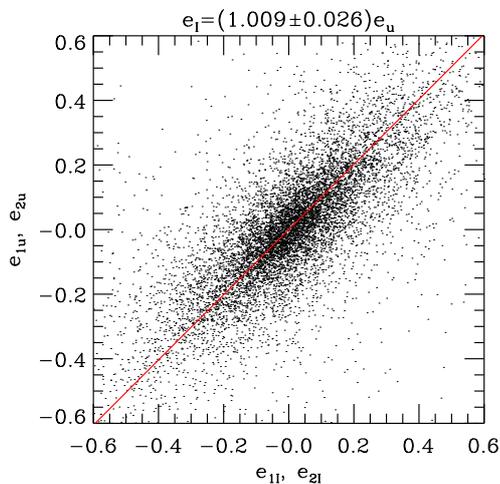}
  \caption{\small Corrected ellipticities of galaxies measured in Megacam-$u^*$
    and CFH12k-$I$ bands in the inner parts of the field. A close consistancy is
    found with $e_I= (1.009 \pm 0.026) e_{u^*}$ with a $\sim0.77$ correlation
    coefficient.}\label{fig:eIeu}
\end{figure}

\subsection{Shear measurement}\label{ssec:wl}
In \citet{adami09}, we presented measurements of the systematic 
orientation of faint Coma cluster galaxies. The small size of these
galaxies required a careful deconvolution of the Point Spread Function
(PSF) smearing effect which was done with the same procedure we use
here. It is based on the KSB technique \citep{KSB95} and the
pipeline has been applied to ground-based imaging from CFH12k and
Megacam cameras \citep[\eg][]{gavazzi04,gavazzi07a}.
We used the $u^*$ channel to measure shapes because it is the
one with largest spatial coverage, and also the most homogeneous since
the images in the other bands are built from two adjacent pointings. The
median $u^*$ band seeing FWHM is $1\farcs02$. Despite this relatively
poor value, we were able to successfully correct for PSF
smearing.

The quality of PSF anisotropy corrections can be seen in the left panels
of Fig.~1 in \citet{adami09}.
The good overall image quality is reflected in the raw and corrected 
ellipticities of stars. The low scatter in the corrected ellipticities,
$0.0028$ rms, is also a good assessment of the controled systematics.

Based on mock images designed to simulate Megacam data for the CFHTLS survey
and kindly made available to us by B.~Rowe, we were able to assess the
reliability of our shear measurements, especially for the isotropic part of
the PSF smearing effect. These simulations are similar to
STEP2 simulations \citep{massey07step} as they capture the complex shape of
sources by the use of shapelet models. Following notations of the STEP
shear measurement project \citep{heymans06step,massey07step}, residual
additive calibration errors are found to be $c\le 0.003$ and multiplicative
$m \le 0.05$, which is significantly smaller than the statistical errors
present in our data (see below).

Because redder bands are usually used for weak lensing studies, we
checked that PSF-corrected ellipticities of central objects measured in 
both Megacam-$u^*$ and CFH12k-$I$ bands are consistent. This is found to be
the case with great accuracy as shown in Fig.~\ref{fig:eIeu}. We checked
that the ellipticity difference $e_I -e_{u^*}$ (for both components 1 and 2
taken together or considered individually) does not correlate with either
$e_{u^*}$, $u^*$ magnitude, $I$ magnitude  or $r_{h,u^*}$ the $u^*$-band
half light radius.

\begin{figure}[htb]
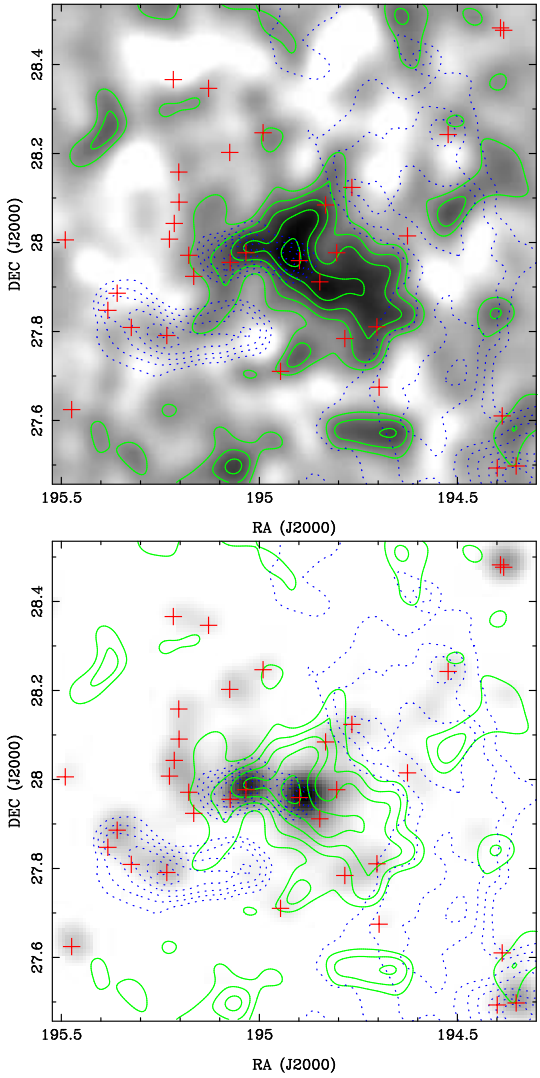

  \centering
  \includegraphics[width=7cm]{kappa+cmemb.eps}
  \includegraphics[width=7cm]{light+cmemb+Xneumann.eps}
  \caption{{\it
  Top panel:} Convergence map for the Coma cluster. Green contours
  represent signal-to-noise ratios of $1,\,2,\ldots5$, corresponding to
  $\kappa=0.01,\,0.02,\ldots0.05$. Red crosses represent the bright
  cluster members lying on the red sequence.\ \ \ {\it Bottom panel:}
  Gray-scale view of the luminosity distribution of cluster red sequence
  members with the overlaid contours in green. The blue dotted contours
  show the excess X-ray emission over a smooth $\beta$-model X-ray
  emissivity map \citep{neumann03}.}\label{kappamap}
\end{figure}

\subsection{Convergence map}\label{ssec:map}
From the source catalogue, we can infer the shear field $\gamma(\vec{\theta})$
and deduce the associated convergence field $\kappa(\vec{\theta})$ which is
the  projected surface mass density expressed in units of the 
critical density $\scrit= c^2 (4\pi G)^{-1} D_{\rm S}/(D_{\rm L} D_{\rm LS})$.
Here $D_{\rm S}/D_{\rm LS}$ is replaced by the source population average
inverse $1/\overline{w}$ which leads to a critical density
$\scrit=1.69 \times 10^{10} \h \msun/\kpc^2$.

 Shear and convergence are related by:
\begin{equation}\label{eq:KS93}
  \kappa(\vec{\theta}) = \int_{\mathbb{R}^2} K(\vec{\theta}-\vec{\vartheta})^* 
  \gamma(\vec{\vartheta}) \der^2 \vec{\vartheta},
\end{equation}
$K(\vec{\theta})=\frac{1}{\pi}\frac{-1}{(\theta_1-i \theta_2)^2}$
is a complex convolution kernel \citep{kaiser93}. The shear field is smoothed
with a Gaussian $G(\theta)\propto\exp(-\frac{\theta^2}{2\theta_s^2})$
with $\theta_s=170\arcsec\simeq80\hmkpc$. The convergence field is
smoothed by the same filter. The convergence map presents correlated
noise properties \citep{waerbeke00} characterised by
$\nbg\simeq11\,{\rm arcmin}^{-2}$, the number density of background
sources. By bootstrapping the orientation of
background sources, we estimated this noise level to be
$\sigma_{\kappa}=0.0134$ with little variation over the field (except near
edges and masked bright stars).

The top panel of Fig.~\ref{kappamap} shows the reconstructed convergence map
$1\times1\, {\rm deg}^2$ field of view with contours
showing a $> 5\sigma$ significance central peak along with various
substructures. The high physical resolution of
$\sim 80 \hmkpc$ that is made possible by the vicinity of Coma somewhat
balances the low associated lensing efficiency.
The position of the brightest cluster galaxies with $r\le 14$ is shown
as red crosses. For comparison we also show in the lower
panel of Fig.~\ref{kappamap} the luminosity-weighted r-band luminosity
distribution coming from bright $r\le 19$ cluster members that
lie on the Red Sequence defined by the relation
$0.72 \le (g-r)+(r-14)*0.035 \le 0.87 $. The convergence
 contours are overlaid.
For this latter study we used the more extended SDSS DR7 photometry,
which is better suited for investigation of the bright end of the luminosity
function\footnote{\url{http://cas.sdss.org/astro/en/tools/}}. This panel also
shows contours of X-ray emissivity in excess of a smooth
$\beta$-model profile \citep{neumann03}.

\section{Radial mass profile}\label{sec:profile}
We now investigate the azimuthally-averaged tangential
shear profile $\gamma_t(R)$ which is simply related to the 
azimuthally-averaged projected 
surface mass density profile $\Sigma(R)$ by the relation
\begin{equation}\label{eq:shear1d}
  \scrit \gamma_t(R) \equiv \Delta \Sigma(R) = M(<R)/(\pi R^2) - \Sigma(R)\,,
\end{equation}
where we have defined the frequently used rescaled shear $\Delta \Sigma$
and $M(<R)$ is the projected mass enclosed by radius $R$.

\begin{figure}[htb]
  \centering
  \includegraphics[width=8.5cm,height=6cm]{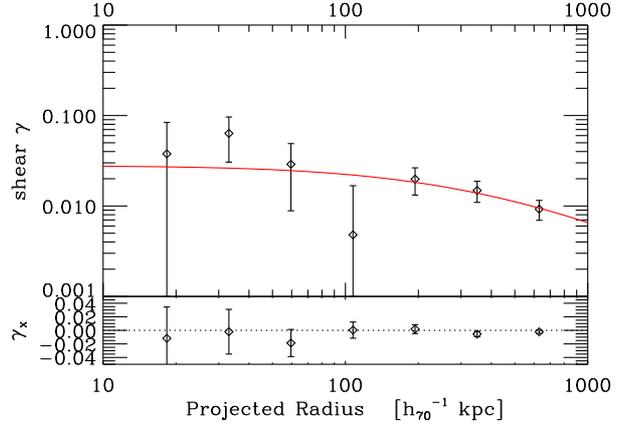}
  \caption{{\it Top panel:} Radial shear profile measured in Coma as a function of projected distance. The best-fit NFW profile is overlaid.\ \ \ {\it Bottom panel:} B-mode shear profile showing the negligible residuals in the rotated shear component.}\label{profile}
\end{figure}

We measure the radial shear profile average in circular annuli centered
on the peak of the convergence map ($\alpha$=12:59:39.007,
$\delta$=+27:57:55.93) which is only 40$\arcsec$ east of NGC~4874,
\ie~in the direction of NGC~4889, which are respectively the brightest
and second brightest member galaxies.
Fig.~\ref{profile} shows the radial shear profile
$\gamma_t(R)$ out to the edge of the Megacam field of view ($\sim$1~Mpc). 
The bottom panel of this figure shows the
same profile once galaxies are rotated by 45$^\circ$, which is 
the curl or B-mode component of the ellipticity field. In the
absence of systematic PSF correction residuals, this rotated shear
profile should be consistent with zero at all scales. This is what we
observe.

\begin{figure}[htb]
  \centering
  \includegraphics[width=\hsize]{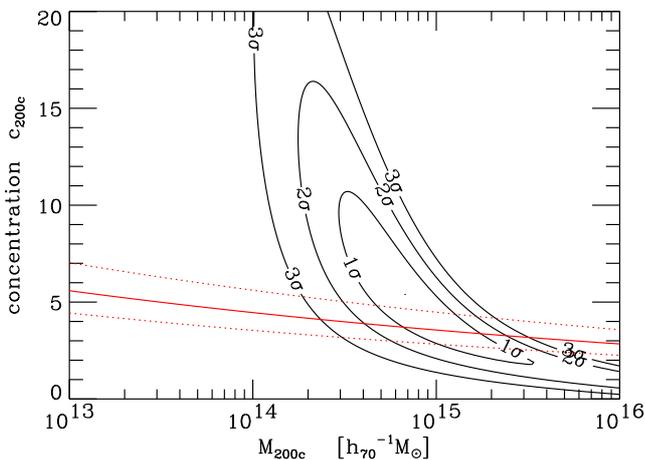}
  \caption{$M_{\rm 200c}$ and $c_{\rm 200c}$
    parameters constraints of the NFW density profile. The $M-c$ relation predicted by
    simulation and WMAP5 cosmology (with its $1\sigma$ scatter) are shown
    in red.}\label{contours}
\end{figure}

We attempted to fit a radial shear profile as predicted by the NFW
mass density profile of the form
\begin{equation}
  \rho(r) = \rho_s (r/r_s)^{-1} \left( 1 + r/r_s\right)^{-2}
\end{equation}
coming from cosmological simulations \citep{NFW97}.  The corresponding
lensing quantities were derived by \citet{bartelmann96}.  The two
relevant quantities that we fit are the mass $M_{\rm 200c}$ enclosed in
the radius $r_{\rm 200c}$ in which the mean density is 200 times the
critical density $\rho_{\rm crit}$ and the concentration parameter
$c_{\rm 200c} = r_{\rm 200c}/r_s$. This implies that
$\rho_s= \rho_{\rm crit} \frac{200}{3} c^3 / \left[ \ln(1+c) - c/(1+c)\right]$.  
Fig.~\ref{contours} shows the constraints we obtain on these
two parameters. Marginal distributions yield the following
constraints\footnote{Assuming a flat uniform prior in $\log M_{\rm
200c}$ betwen 13 and 16 and on
$c_{\rm 200c}$ between 0.01 and 20 to infer respectively the marginal PDFs
$p(c)$ and $p(M_{\rm 200c})$}:
$M_{\rm 200c}= 5.1\mypm{4.3}{2.1} \times 10^{14} \hmmsun$
and $c_{\rm 200c}=5.0\mypm{3.2}{2.5}$, which corresponds to
$r_{\rm 200c}=1.8\mypm{0.6}{0.3}\Mpc$.
Fig.~\ref{contours} also shows the mass-concentration relation and its
$1\sigma$ dispersion that were recently reported by \citet{maccio08}
assuming WMAP5 cosmological parameters \citep{komatsu08}.  The
two-dimensional contours are in good agreement with these predictions.
Taking the conditional $p(c_{\rm 200c}\vert M_{\rm 200c})$ from
\citet{maccio08} as a prior on $c_{\rm 200c}$, we marginalize again over
the poorly constrained concentration parameter and obtain
constraints on the mass
$M^\prime_{\rm 200c}= 9.7\mypm{6.1}{3.5} \times 10^{14} \hmmsun$
or again $r_{\rm 200c}=2.2\mypm{0.3}{0.2}\Mpc$\ .

In order to allow a comparison with other mass estimates
in the literature, we calculate the virial mass $M_{\rm vir}\equiv
M_{\rm 100c}$ and the corresponding concentration $c_{\rm vir}\equiv c_{\rm
100c}$, since a density contrast $\Delta_{\rm vir}\simeq100$ is better
suited for the assumed cosmology. All our results regarding
both $(M_{\rm 200c},c_{\rm 200c})$ and $(M_{\rm vir},c_{\rm vir})$ as well as
corresponding $r_{\rm 200c}$ and $r_{\rm vir}$ values  are
given in Table~\ref{tab:massvalues}

\begin{table}
  \caption{Summary of mass estimates with and without priors on the mass-concentration relation \citep{maccio08}.}
  \label{tab:massvalues}
  \centering
  \begin{tabular}[h!]{lcc}
    \hline\hline
         & no prior  & prior $p(c\vert M)$ \\
    \hline
    $M_{\rm 200c}/10^{14} \hmmsun$ & $5.1\mypm{4.3}{2.1}$ & $ 9.7\mypm{6.1}{3.5}$\\
    $c_{\rm 200c}$                 & $5.0\mypm{3.2}{2.5}$ & $3.5\mypm{1.1}{0.9}$ \\
    $r_{\rm 200c}/\hmMpc$          & $1.8\mypm{0.6}{0.3}$ & $2.2\mypm{0.3}{0.2}$ \\
    \hline
    $M_{\rm vir}/10^{14} \hmmsun$ & $6.1\mypm{12.1}{3.5}$ & $11.1\mypm{16.7}{6.1}$\\
    $c_{\rm vir}$                 & $6.7\mypm{4.1}{3.3}$  & $4.9\mypm{1.7}{1.4}$\\
    $r_{\rm vir}/\hmMpc$          & $2.5\mypm{0.8}{0.5}$  & $3.6\mypm{1.1}{0.7}$ \\
    \hline
  \end{tabular}
\end{table}

These results are in good agreement with mass estimates in the literature.
\citet{kubo07} performed a weak lensing mass estimate of Coma based
on the much shallower SDSS data -- resulting in
$\nbg \simeq 1 {\rm arcmin}^{-2}$ -- but extending out to $\sim 14\hmMpc$.
They report a mass $M_{\rm 200c}= 2.7\mypm{3.6}{1.9} \times 10^{15}\hmmsun$
and concentration index $c_{\rm 200c}=3.8\mypm{13.1}{1.8}$
that are statistically consistent with our estimates.
In addition, an earlier X-ray study by \citet{hughes89} gave
$M_{\rm vir}=13\pm2 \times 10^{14} \hmmsun$, and is consistent
with our estimate. Mass measurements coming from galaxy kinematics
are also in good agreement with our mass constraints
\citep[\eg~][]{the86,geller99,lokas03}. Note however that these latter
authors report a concentration parameter $c_{\rm vir}\simeq 9$,
slightly larger that our best-fit value though statistically consistent.

\section{Discussion and conclusion}\label{sec:conclu}

We have performed a weak lensing analysis of the Coma cluster, based on
a deep $u^*$ exposure and on images in other bands which 
allowed to derive photometric redshifts in order to remove as many foreground galaxies
as possible from the background population. We thus obtain a map of the mass distribution
in Coma, the radial shear profile, as well as the mass and concentration
parameter. 

The peak of the convergence map is nearly
coincident with the X-ray center \citep{neumann03} or the
brightest galaxy NGC~4874. The difference is $\sim 27 \hmkpc$, below the
spatial resolution of the convergence map.
We can also notice the correspondence between the
($\alpha$$\sim$194.4, $\delta$$\sim$27.82) 2$\sigma$ detection and the
G12/G14 groups of \citet{adami05} and between the ($\alpha$$\sim$194.5,
$\delta$$\sim$28.03) 2$\sigma$ detection and the northern part of the
west X-ray substructure \citep{neumann03}. Finally, the
convergence map shows a 4$\sigma$ ($\alpha$$\sim$195, $\delta$$\sim$28)
extension overlapping the position of NGC~4889, confirming the
existence of a mass concentration around this galaxy.
Other substructures visible in X-rays and in the optical (e.g. the
NGC~4911 group) do not show up in the convergence map,
indicating that we are dealing with low mass systems.

   The south-west extension of the convergence map is more puzzling. It
does not match the light distribution. The groups G8 and G9
from \citet{adami05} explain part
of this extension, but are not luminous enough to fully account for this
mass concentration. Another explanation would be massive low redshift
structures (in order to still efficiently induce a weak lensing signal
and to be spread largely enough over the sky) but that would not belong
to Coma. The SDSS data used by \citet{gott05} exhibit such a $z \sim 0.08$
``Great Wall'', probably not luminous enough though to explain the south west
mass concentration. Moreover, the magnitude limit of the SDSS
spectroscopy is not very deep, leaving room for other possible large
scale structures not appearing in the SDSS sample. This will have to be
confirmed by a more detailed knowledge of the immediate background of
the Coma cluster (Adami et al. in preparation). 



\begin{acknowledgements}
RG acknowledges Barnaby Rowe for providing mock realistic Megacam images for
shear calibration assessments and T. Futamase and N. Okabe for interesting
discussion about weak lensing in Coma.
We are grateful to the CFHT and Terapix teams, and to
the French CNRS/PNG for financial support. MPU
also acknowledges support from NASA Illinois space grant
NGT5-40073 and from Northwestern University. 
\end{acknowledgements}

\bibliographystyle{aa}
\bibliography{references}


\end{document}